\documentclass[9pt,twocolumn,twoside]{osajnl}

\journal{ol} 

\setboolean{shortarticle}{true}

\title{On a universal solution to the transport-of-intensity equation}
\author[1,3,4]{Jialin Zhang}
\author[1,3,5]{Qian Chen}
\author[1,3,4]{Jiasong Sun}
\author[2,6]{Long Tian}
\author[1,3,4,7]{Chao Zuo}

\affil[1]{School of Electronic and Optical Engineering, Nanjing University of Science and Technology, No. 200 Xiaolingwei Street, Nanjing, Jiangsu 210094, China}
\affil[2]{School of Science, Nanjing University of Science and Technology, No. 200 Xiaolingwei Street, Nanjing, Jiangsu 210094, China}
\affil[3]{Jiangsu Key Laboratory of Spectral Imaging \& Intelligent Sense, Nanjing, Jiangsu 210094, China}
\affil[4]{Smart Computational Imaging Laboratory (SCILab), Nanjing University of Science and Technology, Nanjing, Jiangsu Province 210094, China}
\affil[5]{chenqian@njust.edu.cn}
\affil[6]{tianlong19850812@163.com}
\affil[7]{zuochao@njust.edu.cn}

\begin{abstract}

Transport-of-intensity equation (TIE) is one of the most well-known approaches for phase retrieval and quantitative phase imaging.
It directly recovers the quantitative phase distribution of an optical field by through-focus intensity measurements in a noninterferometic, deterministic manner.
Nevertheless, the accuracy and validity of state-of-the-art TIE solvers depend on restrictive preknowledge or assumptions, including appropriate boundary conditions, a well-defined closed region, and quasi-uniform in-focus intensity distribution, which, however, cannot be strictly satisfied simultaneously under practical experimental conditions.
In this Letter, we propose a universal solution to TIE with the advantages of high accuracy, convergence guarantee, applicability to arbitrarily-shaped regions, and simplified implementation and computation.
With the ``maximum intensity assumption'', we firstly simplified TIE as a standard Possion equation to get an initial guess of the solution. Then the initial solution is further refined iteratively by solving the same Possion equation, and thus, the instability associated with the division by zero/small intensity values and large intensity variations can be effectively bypassed. Simulations and experiments with arbitrary phase, arbitrary aperture shapes, and nonuniform intensity distributions verify the effectiveness and universality of the proposed method.
\end{abstract}

\setboolean{displaycopyright}{true}

\begin{document}
\maketitle
\noindent
Transport of intensity equation (TIE) is a powerful tool for phase retrieval and quantitative phase imaging (QPI) \cite{teague1983deterministic}. Over past decades, this method has attracted numerous attentions due to its unique advantages over interferometric approaches, such as being deterministic, non-interferometric \cite{paganin1998noninterferometric}, applicable to temporally/spatially coherent beams (e.g., LED or halogen lamp), phase-unwrapping-free, and stable to the environmental disturbances. It has been widely used over a wide range of light- and electron-beams in numerous applications, e.g., X-ray diffraction \cite{nugent1996quantitative}, transmission electron microscopy \cite{bajt2000quantitative}, neutron radiography \cite{allman2000imaging}, and optical quantitative phase imaging \cite{barty1998quantitative}.

TIE is a second-order elliptic partial differential equation that describes the quantitative relationship between the quantitative phase and the intensity variation along the propagation direction. The ``well-posedness'' and ``uniqueness'' of the solution require a strictly positive intensity and, more importantly, the precise knowledge of (Dirichlet, Neumann) boundary conditions \cite{gureyev1995partially}, which, however, are difficult to measure or to know as a priori. For example, to obtain the Dirichlet boundary condition, one needs to know the phase values at the region boundary \cite{teague1983deterministic} or manually select the ``smooth region'' inside the phase distribution \cite{parvizi2015practical}. The fast Fourier transform (FFT)-based solver \cite{paganin1998noninterferometric} (FFT-TIE solver) can be used to avoid the complexity of obtaining such boundary conditions, but it assumes that the finite signal is periodic and repetitive due to the cyclic nature of the discrete Fourier transform. Nevertheless, this situation is rather restrictive and does not reflect general experimental conditions. When the actual experimental situation violates those imposed assumptions, e.g., objects located at the image borders, severe boundary artifacts will appear, seriously affecting the accuracy of the phase reconstruction \cite{zuo2014boundary}. On the other hand, for certain phase functions (such as tilt, defocus, and astigmatism), the defocus-induced intensity derivative signals are all concentrated at the boundary region. If the boundary conditions are not considered, the phase can never be recovered correctly.


The key to solving the above-mentioned issues is obtaining the boundary signals, especially for experiments. Roddier \emph{et al.} \cite{roddier1988curvature2} successfully obtained the boundary values from the intensity measurements at the pupil boundary with the assumption of uniform in-focus intensity distribution in adaptive optics. It is still necessary but difficult to distinguish the boundary signals from the interior intensity derivative. Moreover, this assumption is difficult to satisfy in the field of QPI, especially for objects with strong absorption. To solve these problems, Zuo \emph{et al.} \cite{zuo2014boundary} found that around the introduced aperture edge, the inhomogeneous Neumann boundary conditions can be directly obtained, and then TIE can be effectively and efficiently solved using the fast discrete cosine transform (DCT) with a rectangular aperture. But this fast solution is only available for a rectangular aperture because the DCT only applies to rectangular domains. In practice, it is quite challenging to add an exactly-rectangular aperture since the difficulties in aperture fabrication and system alignment, or the other existing pupils (e.g., reflecting telescopes) may obstruct the system aperture to be rectangular \cite{huang2015phase}. Besides the difficulties in obtaining boundary conditions, the phase discrepancy resulting from Teague's assumption is also another notorious problem for the phase retrieval based on TIE \cite{schmalz2011phase,ferrari2014transport}. Though several iterative TIE algorithms have been proposed to compensate for the phase discrepancy, and they do work under certain conditions \cite{zuo2014phase,huang2015phase}, there is no theoretical guarantee for the convergence. When significant intensity variations or intensity singularities (small intensity values) exist, the iterations will become unstable and prone to divergence \cite{zuo2014phase}.

From the above, we know that the accuracy and validity of state-of-the-art TIE solvers depends on restrictive preknowledge or assumptions, including appropriate boundary conditions, a well-defined closed region, and quasi-uniform in-focus intensity distribution, which, however, cannot be strictly satisfied simultaneously under practical experimental conditions. Ideally, there are at least four issues need to be addressed for a desired TIE solver. 1) It should account for inhomogeneous boundary conditions with experimentally measured boundary signal. 2) It should be applicable to arbitrarily-shaped apertures with arbitrarily-distributed intensity function (hard/soft aperture, large intensity variations, and small intensity values). 3) It provides accurate solution without phase discrepancy. 4) It should be efficient and strictly convergent (if it is iterative). Based on these considerations, we propose a universal solution to TIE (US-TIE) with the advantages of high-accuracy, convergence guarantee, applicability to arbitrarily-shaped regions, and simplified implementation and computation. Let us firstly start with TIE originally proposed by Teague \cite{teague1983deterministic}:
\begin{equation}
- k\frac{{\partial I\left( {\mathbf{r}} \right)}}{{\partial z}} = \nabla  \cdot \left[ {I\left( {\mathbf{r}} \right)\nabla \phi \left( {\mathbf{r}} \right)} \right],
\label{eq:TIE}
\end{equation}
where $I\left( {\mathbf{r}} \right)$ is the in-focus intensity, $\mathbf{r}$ is the 2D spatial coordinates, $\phi \left( {\mathbf{r}} \right)$ is the phase distribution to be solved, $k$ is the wave number. To simplify Eq. \ref{eq:TIE}, Teague \cite{teague1983deterministic} suggested to introduce an auxiliary function $\psi \left( {\bf{r}} \right)$ such that $\nabla \psi \left( {\bf{r}} \right) = I\left( {\bf{r}} \right)\nabla \phi \left( {\bf{r}} \right)$.
Then Eq. \ref{eq:TIE} can be simplified into two standard Poisson equations:
\begin{equation}
{\kern 15pt}  - k\frac{{\partial I\left( {\bf{r}} \right)}}{{\partial z}} = {\nabla ^{\rm{2}}}\psi \left( {\bf{r}} \right)\\
\label{eq:Teague1}
\end{equation}
\begin{equation}
\nabla  \cdot \left[ {I{{\left( {\bf{r}} \right)}^{{\rm{ - 1}}}}\nabla \psi \left( {\bf{r}} \right)} \right]{\rm{ = }}{\nabla ^{\rm{2}}}\phi \left( {\bf{r}} \right),
\label{eq:Teague2}
\end{equation}
which can be solved with use of FFT \cite{paganin1998noninterferometric} or DCT \cite{zuo2014boundary} efficiently. The solution can be simply denoted as:
\begin{equation}
\phi \left( {\mathbf{r}} \right) =  - k{\nabla ^{ - 2}}\left\{ {\nabla \cdot \left\{ {\frac{1}{{I\left( {\mathbf{r}} \right)}}\nabla {\nabla ^{ - 2}}\left[ {\frac{{\partial I\left( {\mathbf{r}} \right)}}{{\partial z}}} \right]} \right\}} \right\}.
\label{eq:TIE_Phase_old}
\end{equation}
Equation \ref{eq:TIE_Phase_old} is usually referred to the ``Teague's assumption'', which suggests that the transverse flux is conservative so that can be fully characterized by a scalar potential \cite{teague1983deterministic}. However, there is no guarantee that the transverse flux is always conservative due to the curl component in the Helmholtz decomposition, especially when large intensity variations or singularities exist, resulting in non-ignorable phase discrepancy \cite{schmalz2011phase,zuo2014phase}. Moreover, in experimental conditions, the intensity captured at the in-focus plane ${I\left( {\mathbf{r}} \right)}$ often contains dark regions where the intensity value approaches or equals zero, especially for the introduction of the arbitrarily-shaped aperture. This precludes the direct use of Eq. \ref{eq:TIE_Phase_old} for phase reconstruction since ${I\left( {\mathbf{r}} \right)}$ appears in the denominator. In fact, the ``divide-by-zero'' is the major factor behind the phase discrepancy and the instability of the state-of-the-art TIE solvers. Fortunately, we found that when the in-focus intensity is uniform ($\tilde I$), it can be directly pulled out from the gradient operator, and TIE directly boils down to a Poisson equation $- k\frac{{\partial I\left( {\bf{r}} \right)}}{{\partial z}} = \tilde I{\nabla ^2}\phi \left( {\bf{r}} \right)$.
\begin{figure}[!b]
\centering
\includegraphics[width=\linewidth]{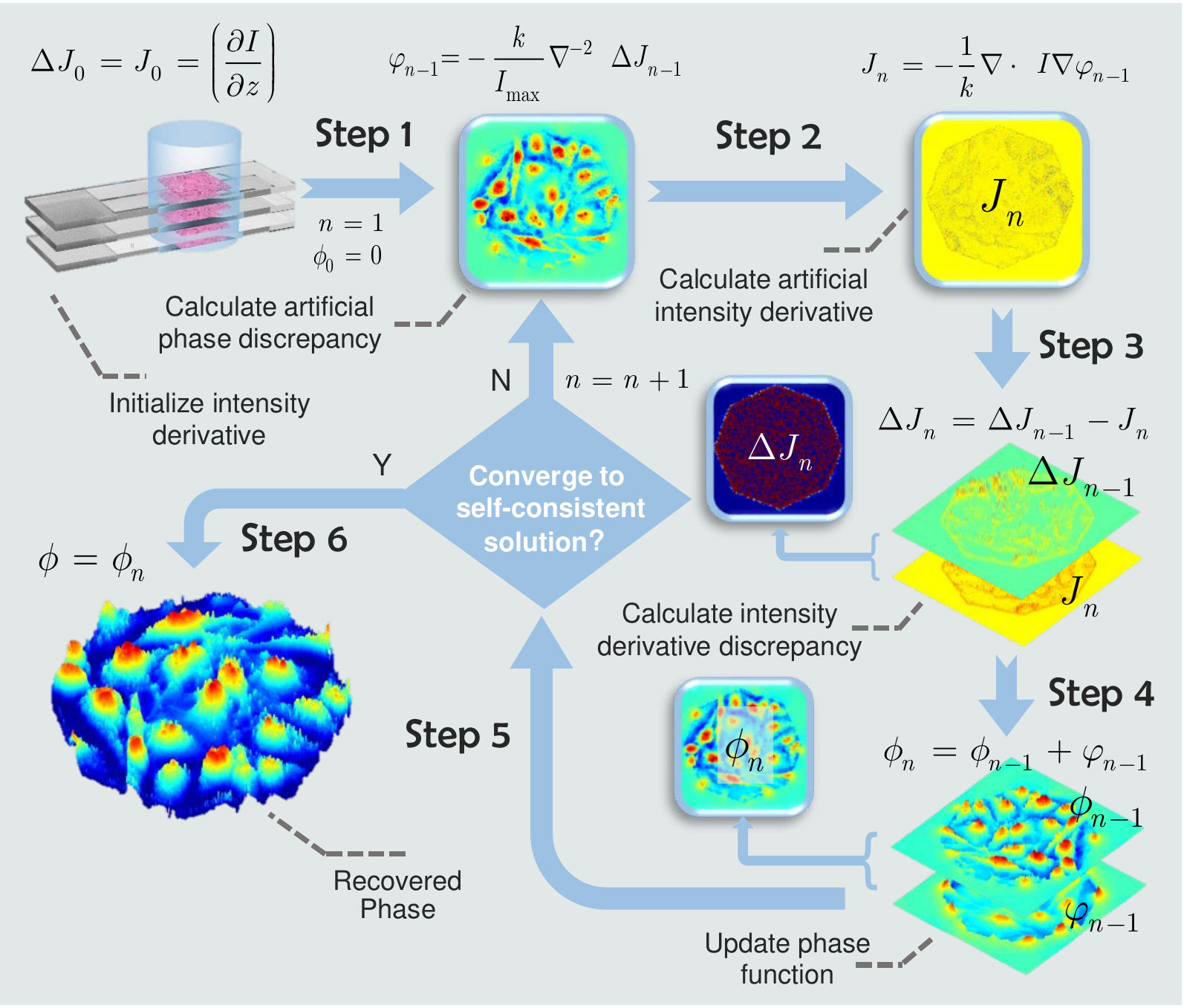}
\caption{The flow chart of the phase retrieval based on the US-TIE method.}
\label{fig:fig1}
\end{figure}
In such case, the transverse flux is always conservative and Teague's assumption will be no longer necessary, and thus, the solution to TIE becomes trivial and no phase discrepancy will be induced. Inspired by this observation, when solving TIE, we first assume that the in-focus image intensity is uniform with a constant value $I_{max}$, where $I_{max}$ is  the maximum value of the in-focus image intensity, to bypass the difficulties associated with the ``divide-by-zero'' problem. Then, the solution of the phase takes the following form:
\begin{equation}
\phi \left( {\mathbf{r}} \right) = -\frac{{k}}{I_{max}}{\nabla ^{-2}}\left[ {\frac{{\partial I\left( {\mathbf{r}} \right)}}{{\partial z}}} \right],
\label{eq:TIE_Phase_New}
\end{equation}
where the inverse Laplacian operator ${\nabla ^{-2}}$ can be effectively implemented by only one pair of FFT. But it should be noted that the actual intensity distribution can be an arbitrary function, so the maximum intensity assumption $I_{max}$ used here is obviously ``unreasonable'', and the phase calculated by Eq. \ref{eq:TIE_Phase_New} is usually inaccurate. Thus, the next step is to treat the inaccurate phase $\varphi _0\left( {\mathbf{r}} \right)$ as an initial solution, and substitute it back to Eq. \ref{eq:TIE}. The inaccurate solution results in inconsistency between the calculated intensity derivative $J_n$ and the real measurement value, which is treated as the error signal for another round of phase reconstruction. The solution $\varphi _0\left( {\mathbf{r}} \right)$ is also taken as the ``correction term'', which is added back to $\phi_0\left( {\mathbf{r}} \right)$ to get an updated phase estimate. This completes one iteration of the reconstruction algorithm. The procedure is iteratively repeated until convergence. It should be noted that during the iterative process, the maximum intensity assumption $I_{max}$ is always assumed so that the solution in each iteration can be effectively implemented by two simple FFTs (instead of eight as in conventional FFT-TIE solver). Though the maximum intensity assumption $I_{max}$ is not physically grounded, we access the correctness of our current estimate with the original TIE, which guarantees that when the iterative algorithm converges with a insignificant error signal, the accurate solution to TIE is certain to arrive.

\begin{figure}[!b]
\centering
\includegraphics[width=\linewidth]{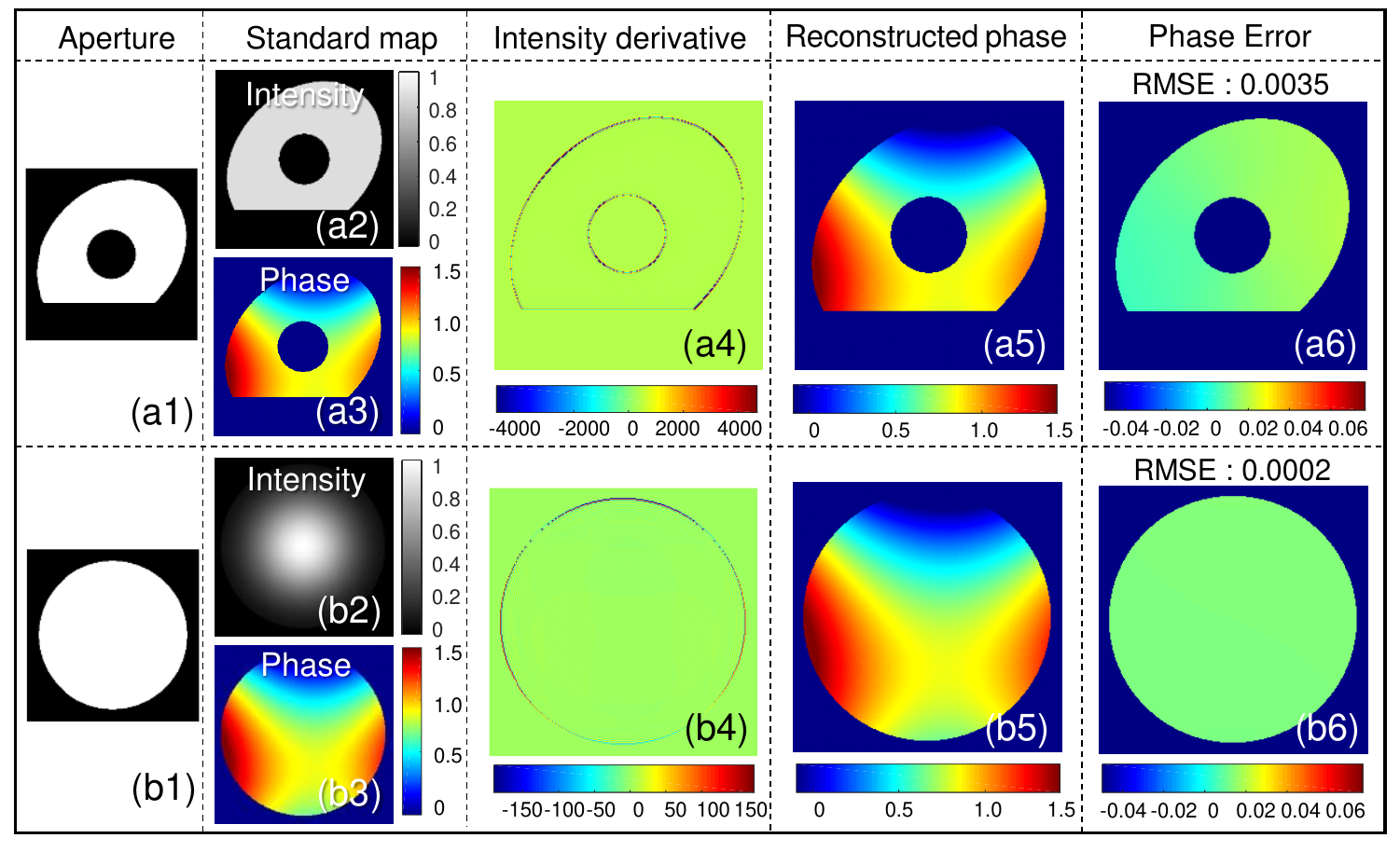}
\caption{Simulation results of US-TIE with irregular and soft-edged apertures. (a1,b1) The shapes of apertures. (a2,b2) In-focus intensities. (a3,b3) True phases. (a4,b4) Intensity derivatives. (a5,b5) Retrieved phases. (a6,b6) Phase errors.}
\label{fig:fig2}
\end{figure}

It should be mentioned that the universal solution proposed here is quite similar to the previous iterative algorithms for compensating the ``phase discrepancy'' owing to the ``Teague's assumption'' \cite{zuo2014phase}, and solving the boundary condition problem with an arbitrarily-shaped aperture  \cite{huang2015phase}.
The only difference is that the maximum intensity assumption $I_{max}$ is introduced here to simplify the solution and prevent numerical instability. More importantly, the maximum intensity assumption $I_{max}$ is also the key to the convergence of the iterative algorithm.
The rigorous proof of convergence of iterative process is given in \emph{\textbf{Supplementary Information}}. In the proof, we firstly establish the Sobolev space ${W^{2,2}}$ priori estimate to the solution on each step \cite{adams2003sobolev}. Then based on the \emph{Sobolev’s Embedding Theorem }\cite{brezis2010functional}, we can derive that the solution on the $n^{th}$ step is comparable to the quantity ${\left( {\frac{{{I_{\max }} - I}}{{{I_{\max }}}}} \right)^n}$. Due to the maximum intensity assumption $I_{\max}$, the common ratio of this geometric sequence can be constrained within the range of 0 to 1, which mathematically, guarantees that the iterative algorithm is convergent.

Simulations are carried out to verify the validity and effectiveness of the proposed method. In the first simulation, an irregular hard aperture with uniform intensity distribution was used, as shown in Figs. \ref{fig:fig2}(a1,a2). The phase distribution is a shifted astigmatism function defined on a ${\text{256}} \times {\text{256}}$ grid: $\phi \left( {\mathbf{r}} \right) = 10r_x^2 - 10r_y^2 - 0.7{r_x} + 2{r_y} + 0.82$, as shown in Figs. \ref{fig:fig2}(a3). Defocused images ($\Delta z = 1\mu m$) are obtained based numerical propagation and the intensity derivative $\partial I/\partial z$ is calculated [Figs. \ref{fig:fig2}(a4)]. Note that for such a phase distribution with zero Laplacian, the intensity derivative only assumes non-zero values at the aperture edge. US-TIE converges after 30 iterations, producing an accurate solution with negligible error (RMSE 0.0035rad),as shown in Figs. \ref{fig:fig2}(a5,a6). The second simulation tests the US-TIE under soft-edged illumination. The intensity profile is a Gaussian beam within a circular aperture [Figs. \ref{fig:fig2}(b1-b3)]. It is quite challenging case for the TIE phase retrieval because the intensity almost decays to zero at the aperture edge. The US-TIE, once again, successfully recovers the correct phase distribution even in the dark region, as shown in Figs. \ref{fig:fig2}(b4-b6).

\begin{figure}[!htb]
\centering
\includegraphics[width=1.0\linewidth]{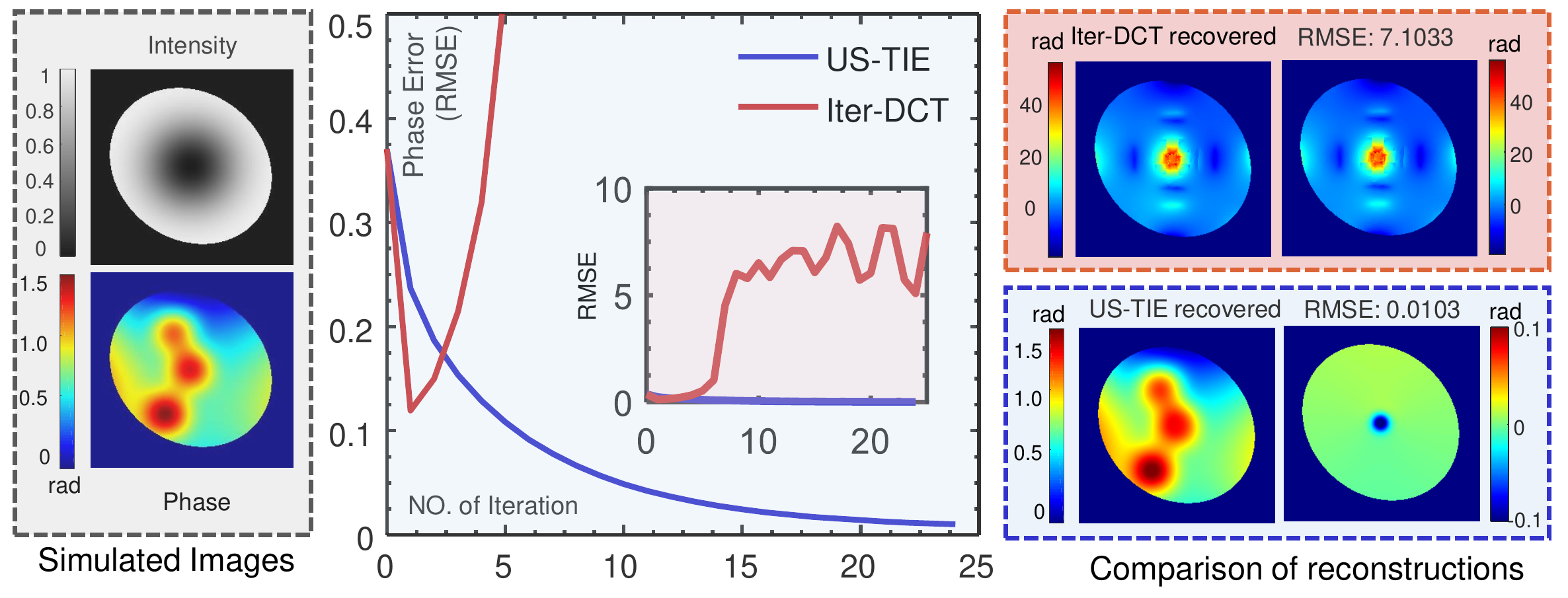}
\caption{Comparison of US-TIE with the iterative DCT (Iter-DCT) method. The left column shows the simulated intensity and phase distribution. The RMSE curves versus the iteration number and reconstruction results are shown in the middle and right colomns, respectively.}
\label{fig:fig3}
\end{figure}

In order to further verify the accuracy and convergence of US-TIE, we compare it with the iterative DCT method (Iter-DCT) \cite{huang2015phase}, which is also applicable to an arbitrarily-shaped aperture. We consider another challenging case that an inverse Gaussian beam (contains an intensity singularity, \textit{i.e.}, zero intensity value), and irregular phase distribution, as shown in the left column of Fig. \ref{fig:fig3}. As the number of iterations increases, the RMSE of US-TIE drops rapidly and finally converges after 29 iterations (RMSE 0.0103rad, total computation time 0.08s with a 3.6 GHz laptop) as shown in Fig. \ref{fig:fig3} (blue curve). Note that phase errors only appear around the intensity singularity, where the phase value is not well defined. In contrast, the RMSE curve of the Iter-DCT method rebounds after one iteration and the iteration diverges (RMSE 7.1033rad) due to the large phase discrepancy resulting from small intensity values (red curve). It should also be mentioned that, in addition to the high accuracy and stable convergence,  US-TIE improves the computational speed ($\sim3.5ms$ per iteration) by more than one order of magnitude compared with the iter-DCT method ($\sim36ms$ per iteration) because only two FFTs are involved for each iteration.

\begin{figure}[!b]
\centering
\includegraphics[width=0.92\linewidth]{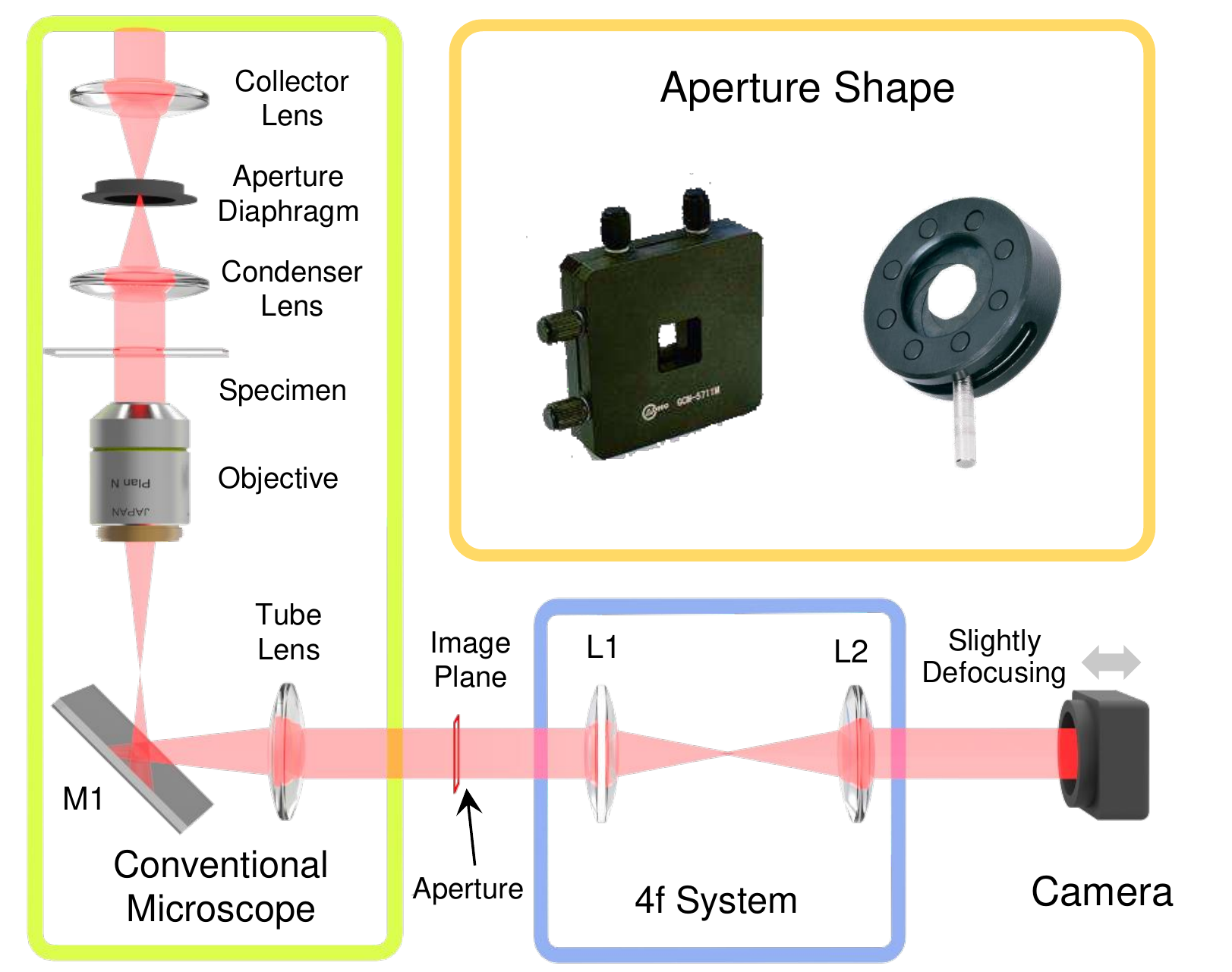}
\caption{The experimental setup is implemented by using an inverted bright-field microscope and a 4\emph{f} system-based TIE module with an aperture at the image plane.}
\label{fig:fig4}
\end{figure}

\begin{figure}[!htb]
\centering
\includegraphics[width=\linewidth]{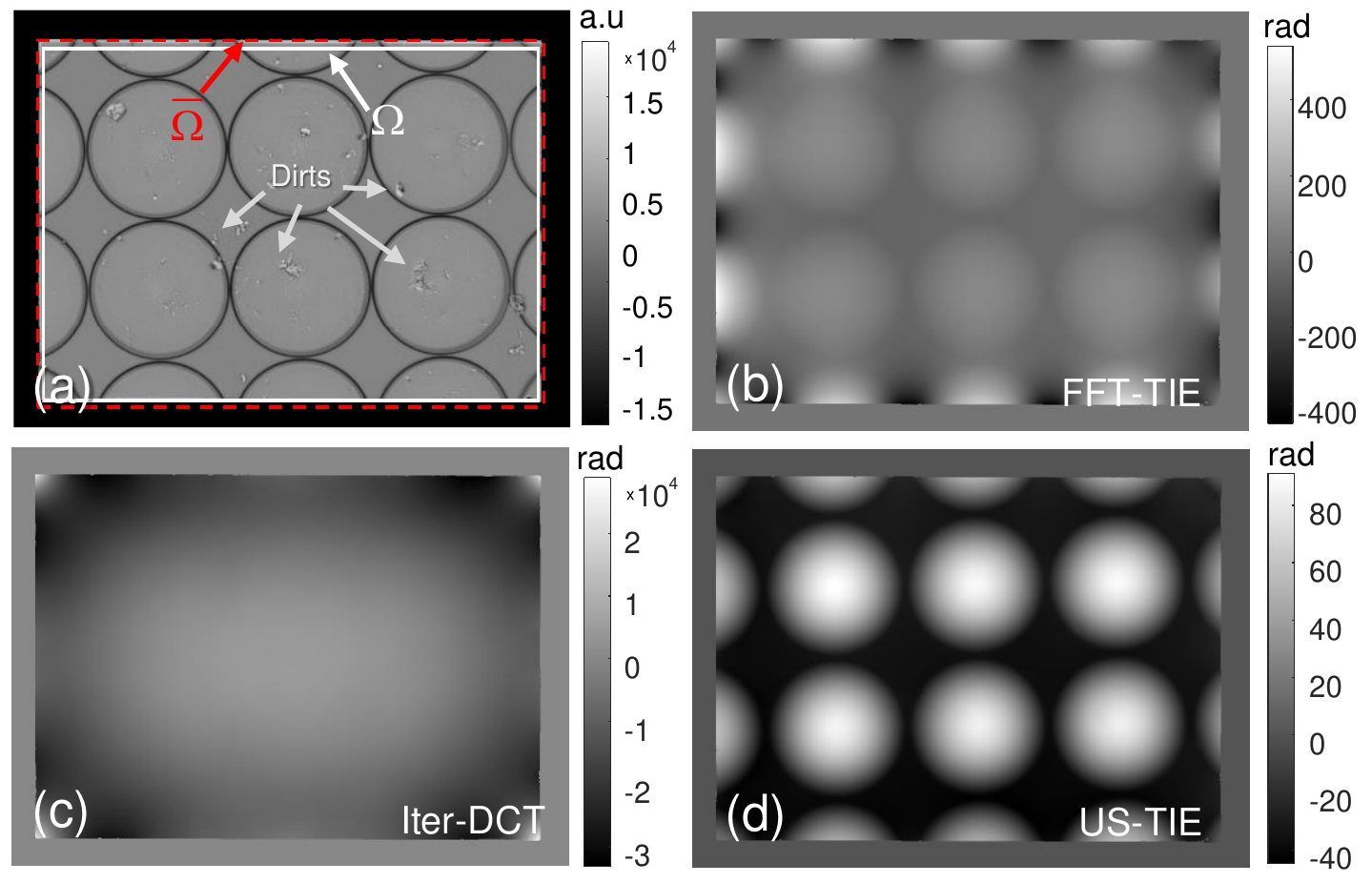}
\caption{(a) The initial intensity derivative. (b) The result reconstructed with the classical FFT-TIE solver. (c-d) The reconstructed phase distribution separately based on the iter-DCT and US-TIE methods.}
\label{fig:fig5}
\end{figure}

Two experiments are performed to demonstrate the practicality of US-TIE. As illustrated in Fig. \ref{fig:fig4}, an inverted bright-field microscope (Olympus IX83) attached with 4\emph{f} imaging system is used to acquire the in- and out of focus intensity images by axially translating the camera. The pixel size of the camera (The Imaging Source DMK 72BUC02, ${\text{1280}} \times {\text{960}}$ ) is $2.2\mu m$, and the central wavelength of the illumination is $550nm$. In order to simplify the implementation, apertures (rectangle-like, octagon-like) are inserted on the intermediate image plane of the microscope instead of on the real object plane, as shown in Fig. \ref{fig:fig4}). Figure \ref{fig:fig5} shows the reconstructed results of the microlens array based on the FFT-TIE [Fig. \ref{fig:fig5}(b)], iter-DCT [Fig. \ref{fig:fig5}(c)] and proposed US-TIE [Fig. \ref{fig:fig5}(d)], respectively. The classical FFT-TIE solver is implemented to retrieve the phase within the rectangular region $\Omega$ only [white rectangle in Fig. \ref{fig:fig5}(a)]. The profiles of microlenses at the boundary were overestimated and distorted due to the inappropriately used periodic boundary conditions, as shown in Fig. \ref{fig:fig5}(b). The iter-DCT, however, produces erroneous phase reconstruction due to the divergence induced by small intensity values (dirts on the microlens and dark background, see Fig. \ref{fig:fig5}(a)), as shown in Fig. \ref{fig:fig5}(c). In contrast, the proposed US-TIE provides the “unbiased” solution of TIE that is free from any boundary errors and phase discrepancies.

\begin{figure}[!b]
\centering
\includegraphics[width=\linewidth]{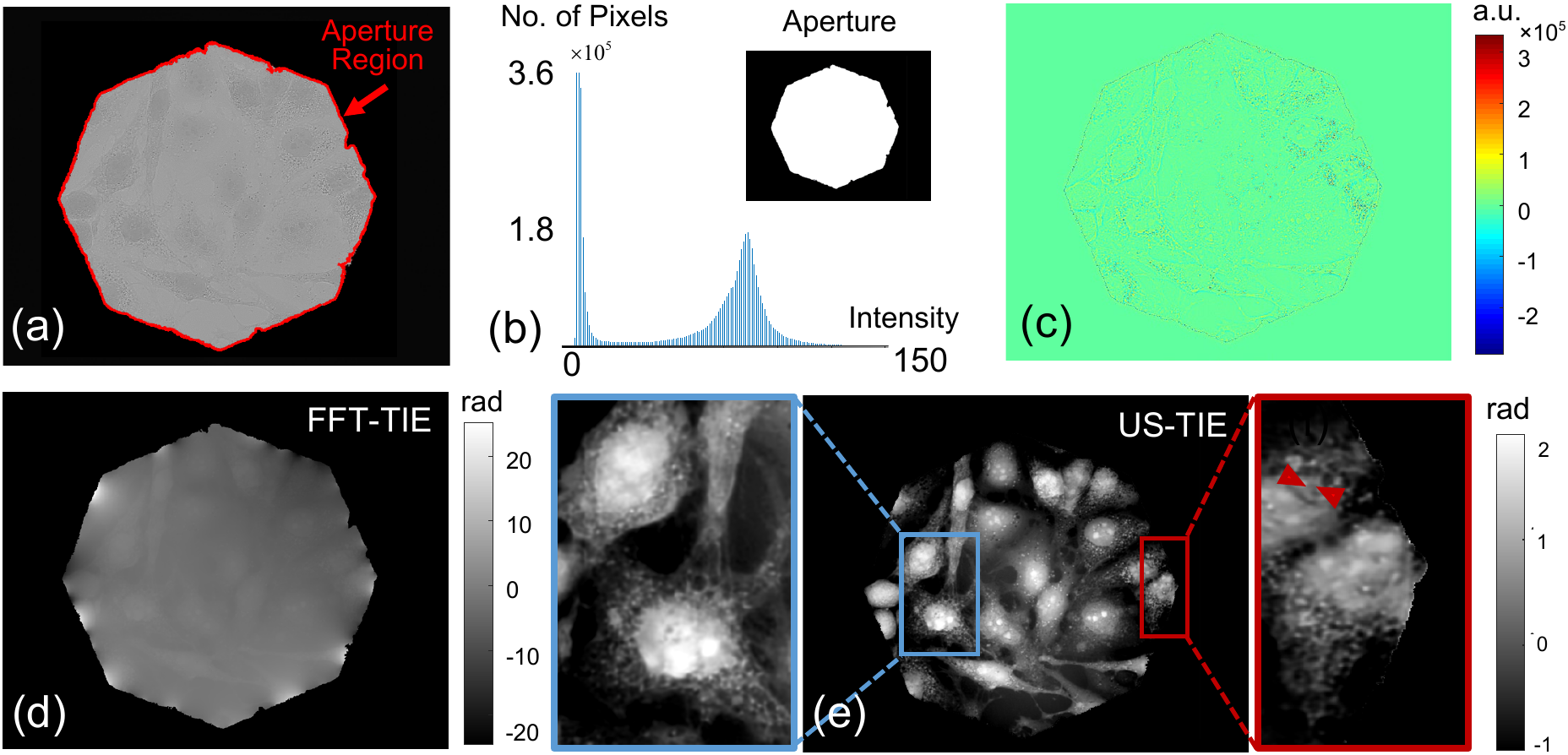}
\caption{Experimentally captured intensity at focus plane (a) with its histogram (b), and the initial intensity derivative (c). The results with the classical FFT-TIE solver (d) and proposed US-TIE method (e), as well as the two zoomed areas within the entire field-of-view.}
\label{fig:fig6}
\end{figure}

Finally, US-TIE is applied to quantitative phase imaging of live HeLa cells with an octagonal aperture. In the raw image shown in Fig. \ref{fig:fig6}(a), it is observed that some cells are located across the aperture boundary. According to the intensity histogram, the aperture can be obtained by simple thresholds, as shown in Fig. \ref{fig:fig6}(b), and the result [Fig. \ref{fig:fig6}(d)] can be recovered based on the FFT-TIE solver when the zero-value region is filled with a small constant (0.01). Though intensity zeros are fixed, FFT-TIE is still quite unstable and creates significant artifacts that degrade the whole phase reconstruction prevailingly. The phase retrieved by US-TIE is shown in Fig. \ref{fig:fig6}(e), which reveals sub-cellular features such as the optically thick nucleus, transported vesicle, and Golgi apparatus. In addition, the phase at the boundary of the aperture also can be accurately recovered without any boundary artifacts perceivable, as shown in the red boxed image of Fig. \ref{fig:fig6}(e).

In conclusion, a universal TIE solver is proposed for phase retrieval under nonuniform illuminations, arbitrarily-shaped aperture, and inhomogeneous boundary conditions. Based on the maximum intensity assumption, TIE is iteratively solved based on the simple FFT-based Possion solver efficiently, and the phase discrepancy problem resulted from the phase discrepancy owing to the Teague's assumption can be effectively addressed. The strength of the proposed method lies in its high accuracy, convergence guarantee, and simple implementation, promoting broader applications of TIE in micro-optics inspection, life sciences, and biophotonics. To aid the reader to better understand the implementation of US-TIE, we have uploaded the \textbf{{MATLAB source code and dataset}} for all the simulations and experiments described in this Letter.

\noindent\textbf{Funding.} National Natural Science Foundation of China (61722506), Leading Technology of Jiangsu Basic Research Plan (BK20192003), Outstanding Youth Foundation of Jiangsu Province (BK20170034), The Key Research and Development Program of Jiangsu Province (BE2017162), J. Zhang acknowledges financial support from China Scholarship Council.
\bibliography{sample}

\bibliographyfullrefs{sample}

\end{document}